\documentclass[aps,twocolumn,floatfix,showpacs]{revtex4}
\usepackage{graphicx,bm}
\usepackage{amsmath}
\usepackage{amssymb}
\begin{document}
\title {Resonant scattering of ultracold atoms in low dimensions}
\author{Ludovic Pricoupenko}
\affiliation
{$^{1}$Laboratoire   de   Physique  Th\'{e}orique de la Mati\`{e}re
Condens\'{e}e, Universit\'{e} Pierre et Marie Curie, case courier 121,
4 place Jussieu, 75252 Paris Cedex  05, France.}
\date{\today}
\begin{abstract}
Low energy scattering amplitudes for two atoms in one- and
two-dimensional atomic wave guides are derived for short range
isotropic and resonant interactions in high partial wave channels.
Taking into account the finite width of the resonance which was neglected 
in previous works is shown to have important implications in the properties of
the confinement induced resonances. For spin  polarized fermions in 
quasi-1D  wave guides it imposes a strong constraint on the atomic density 
for achieving the Fermi Tonks Girardeau gas. For a planar wave guide, 
the charateristics of the 2D induced scattering resonances in $p$- and 
$d$-wave are determined as a function of the 3D scattering parameters
and of the wave guide frequency.
\end{abstract}
%
%
\maketitle

Recent experimental progress in degenerate atomic gases make
possible accurate studies of quasi-one (1D)
\cite{Richard,Bruno,Kinoshita,Parades,Moritz,Gunter} and quasi-two 
dimensionnal (2D) \cite{Rychtarik,Gunter,Jean} configurations. These 
systems are interesting in view of future applications involving 
coherent manipulation of matter waves 
and can be used also for studying generic phenomenona in low dimensions 
\cite{Richard,Jean}. One major interest of atomic gases is the 
precise  knowledge and experimental control of the low energy interatomic 
collisions: the effective two-body interaction can be tuned with a Feshbach 
resonance \cite{Feshbach} by applying an  external magnetic field and/or with a
Confined Induced Resonance (CIR) by varying the  extension of the trap
in the tight transverse direction \cite{Maxim1D,Petrov2D,Blume1D}.    
These techniques open possibilities for achieving new  types of strongly 
correlated quantum systems. For example, thanks to $s$-wave collisional 
properties in quasi-1D wave guides \cite{Maxim1D}, it has been possible 
to observe the so-called Tonks Girardeau (TG) gas \cite{GirardeauTG} where 
1D hard-core bosons can be mapped onto a system of non interacting fermions 
\cite{Kinoshita,Parades}. For quasi-1D spin polarized fermions an analogous 
regime: the Fermionic Tonks Girardeau (FTG) gas is subject to intensive 
studies \cite{Wright,Bender,AnnaFTG}. In this case, as a consequence of the 
Pauli exclusion principle, atoms interact predominantly in the $p$-wave channel 
and it has been predicted that the strongly interacting 1D polarized Fermi gas 
can be mapped onto a non interacting Bose gas.

In this letter, it is shown that the finite width of the high partial waves 
resonances is an essential feature in the properties of the corresponding CIR. 
As a consequence, the resonant fully polarized Fermi atomic gas in quasi-1D 
traps can reach the FTG regime in a very dilute limit only (which means that for 
realistic trap parameters, only few particles systems can undergo this regime) 
and is more generally  described by a narrow BEC-BCS crossover. Violation of 
the fermion-boson mapping theorem \cite{Cheon} opens relevant issues on the 
quasi-1D resonant polarized Fermi gas such as the equation of state and the 
shift and damping of the collective modes which are only known at the FTG 
limit \cite{AnnaFTG}. Motivated by the exciting predictions of exotic superfluid 
phases \cite{Gurarie,Tewari} with non-Abelian statisitics, the $p$-wave scattering 
amplitude in quasi-2D atomic wave guide induced by a resonant interaction is 
derived. The effective range parameter which were neglected in Ref.\cite{Idziaszek} 
appears essential for the  determination of the 2D low energy scattering parameters.
Due to the possibility of achieving $d$-wave resonances \cite{Chin} in 3D systems, 
the quasi-2D $d$-wave scattering amplitude is also derived and the characteristics 
of this new CIR are depicted.

In the following, the true interatomic forces are supposed to be short
range and isotropic. Moreover they are considered in the neighborhood
of a resonance. The 3D low energy collisional properties of two atoms
are then parameterized in each partial wave (${l \geq 0}$) by the following 
phase shift ${\delta_l(q)}$:
\begin{equation}
q^{2l+1} \cot \delta_l(q) = - \frac{1}{w_l} - \alpha_l q^2 + O(q^4) ,
\label{eq:phaseshift}
\end{equation}
in Eq.(\ref{eq:phaseshift}) $q$ is the relative momentum of the two
colliding atoms, and the resonant regime in the $l$-partial wave is
achieved for ${|w_l| \gg R^{2l+1}}$, where $R$ is the characteristic
radius of the pairwise  potential ($w_0$
is the usual scattering length and $w_1$ is the scattering volume).
The effective range parameter $\alpha_l$ is linked to the width of the 
resonance: a large (small) value of ${\alpha_l R^{2l-1}}$ corresponds  to a 
narrow (relatively broad) resonance. In the neighborhood of the 
resonance and for ${l \geq 1}$, it  has been shown recently that for a 
finite range potential \cite{pwave,lwaves,Mattia}:
\begin{equation}
{\alpha_l R^{2l-1} \gtrsim (2l-3)!! (2l-1)!!} \ .
\label{eq:inequality}
\end{equation}
Thus ${\alpha_l |w_l| \gg R^2}$,  and  $-\hbar^2/(2\mu\alpha_lw_l)$ is a low energy 
scale ($\mu$ is the reduced mass) which is for large and positive 
values of  $w_l$, nothing but the shallow bound state energy (and the quasi-bound  
state energy for  large  and  negative $w_l$). Hence, unlike $s$-wave  
broad resonances where ${\alpha_0 \equiv  O(R)}$ can be neglected in the low 
energy  limit, for ${l>0}$ the effective range parameter $\alpha_l$ 
which depends on the specific resonance considered, is an essential parameter
involved in the low energy properties. In experiments the resonant 
regime can be achieved by  using  a Feshbach resonance \cite{Feshbach} on the 
spin degree of  freedom.  An external magnetic field $B$ modifies the detuning 
between an open  and a closed channel of the two-body  system, and in the vicinity 
of the resonance  in the partial wave $l$ where ${B\sim B_0}$,  ${w_l^{-1}\propto-(B-B_0)}$
while $\alpha_l$ is almost constant. 

Due to the short range character of interatomic forces, we use the so-called zero 
range approximation. Hence, the pairwise interaction between particles is replaced by 
a source term of vanishing range ${\epsilon \to 0}$ in the Schr\"{o}dinger's equation 
(it is assumed that the characteristic lengths of the trapping potential are 
$\gg R$). Moreover, the trapping potentials considered in this letter are always 
harmonic, so that the center of mass and relative coordinates are decoupled each 
from the other. For a binary collision in the center of mass frame at energy 
${E=\hbar^2q^2/\mu}$, the Schr\"{o}dinger's equation can be transformed into 
an integral equation for the wave function ${|\Psi\rangle}$:
\begin{equation}
|\Psi\rangle = |\Psi_0 \rangle + \frac{2\pi \hbar^2}{\mu} \!
\sum_{l\geq 0} \! \int \!\! \frac{d^3{\bf k}}{(2\pi)^3} \, 
\frac{k^l \delta_{\epsilon}(k)  \left( {\mathcal R_l} \!\cdot\! {\mathcal S}_{l,{\mathbf k}} \right)}{{\mathcal H}_0 - E - i 0^+}  |{\bf k} \rangle ,
\label{eq:Lippmann}
\end{equation}
where ${\mathcal H}_0$ is the free Hamiltonian which includes the external 
potential, and $|\Psi_0\rangle$  belongs  to the kernel of  ${\mathcal H}_0 - E$.  
In Eq.(\ref{eq:Lippmann}), the source term is introduced in the momentum 
representation: ${\delta_\epsilon(k)=\exp(-k^2\epsilon^2/4)}$ is the Fourier
transform of a normalized Gaussian weight having a vanishing range $\epsilon$
\cite{deltalike}, ${\mathcal  R_l}$ and ${\mathcal S}_{l,{\mathbf k}}$
are Symmetric Trace Free  (STF) tensors of rank  $l$, and the notation
$\left( {\mathcal R_l} \!\cdot\!  {\mathcal  S}_{l,{\bf k}} \right)$ means
a contraction between the  two tensors.  The  tensors ${\mathcal S}_{l,{\mathbf   k}}$ are
eigenfunctions of the momentum operator and they appear in a standard multipolar 
expansion in Cartesians coordinates \cite{lwaves}:  
$ {{\mathcal   S}_{l,{\bf  k}}^{[\alpha\beta\dots]}  =   (-1)^l k^{l+1}
\left( \partial_{k_\alpha}  \partial_{k_\beta} \dots \right)  k^{-1}/(2l\!-\!1)!!}
$, where $\{k_\alpha\}$ are the Cartesian components  of the vector ${\bf k}$
(${\alpha \in \{ x,y,z\}}$). Eq.(\ref{eq:Lippmann}) shows explicitly that the interacting 
wave function $|\Psi \rangle$ is the superposition of a regular solution $|\Psi_0\rangle$ 
and of an irregular part generated by the source term \cite{singular}. The correct asymptotic behavior of the 
wave function (for relative coordinates $|{\mathbf r}| \gg R$) is obtained by the determination of the tensors ${\mathcal R_l}$ 
which  fixes the balance between the regular and irregular solutions. For this purpose,
contact conditions (for ${\mathbf r} \to 0$) are imposed in each partial wave and are 
such that without external potential the phase shifts in a scattering process coincide
{\sl  exactly} with the first two terms in the right hand side of Eq.(\ref{eq:phaseshift})
\cite{lwaves}. In the momentum  representation,  the  contact conditions can be written as:
\begin{equation}
\operatornamewithlimits{Reg}_{\epsilon \to 0} \int \frac{d^3 {\bf k}}{(2\pi)^3} k^l
\langle {\bf k} | \Psi \rangle  {\mathcal S}_{l,{\bf k}}
= - \frac{l !\,{\mathcal R_l}}{a_l(q)(2l+1)!\,!}\,  \label{eq:contact} ,
\end{equation}
where  ${a_l^{-1}(q)=(w_l^{-1}+\alpha_l  q^2)}$. In Eq.(\ref{eq:contact}),
$\operatornamewithlimits{Reg}_{\epsilon \to 0}$ means the regular part of the
integral obtained when the formal range $\epsilon$ is set to zero. This way,
the source term in Eq.(\ref{eq:Lippmann}) is itself a functional of 
$|\Psi\rangle$, and a closed equation for ${\mathcal R}_l$ is obtained by combination
of Eqs.(\ref{eq:Lippmann},\ref{eq:contact}). In the next parts, the
scattering problem for $p$- or $d$-wave channels is solved successively 
in 1D and 2D harmonic wave guides.

{\it Linear  atomic  wave guide.}   --- Two atoms   are confined in 
a two-dimensional harmonic trap while they move freely along the third
direction ($z$). In the center of mass frame, the non interacting
Hamiltonian is:
\begin{equation}
{\mathcal H}_0 = - \frac{\hbar^2}{2\mu} \Delta_{\mathbf r} + \frac{1}{2} \mu \omega_\perp^2 \rho^2  - \hbar\omega_\perp ,
\label{eq:H1D}
\end{equation}
where ${{\mathbf r}=  z\,{\bf \hat{e}}_z  + {\boldsymbol \rho}}$ are  the
relative coordinates. For a scattering process at energy ${E = \hbar^2q^2/2\mu}$, 
the state $|\Psi_0\rangle$ in Eq.(\ref{eq:Lippmann}) is: $\langle{\mathbf k}|\Psi_0\rangle = (2\pi) 
\delta(k_z - q)\, \langle{\mathbf k}_{2D} |\phi_{00} \rangle$, where  ${|\phi_{00}\rangle}$ is the ground state  of
the  2D harmonic oscillator in  Eq.(\ref{eq:H1D}), and ${{\mathbf k} =
{\mathbf k}_{2D} + k_z {\bf  \hat{e}}_z}$. Hereafter, the system is in
the monomode regime (${E <  2 \hbar\omega_\perp}$). At large distances ${|z|\gg a_\perp}$ 
(where ${a_\perp=\sqrt{\hbar/\mu \omega_\perp}\gg R}$), the wave function factorizes as 
${\langle\mathbf r | \Psi \rangle  \simeq  \langle\boldsymbol \rho | \phi_{00}\rangle \psi_{1D}(z)}$, and for ${E>0}$:
\begin{eqnarray}
\psi_{1D}(z) =\exp(iqz) + \exp(iq|z|)\bigg[ f^{\rm even}+ \mbox{sign}(z) f^{\rm odd}\bigg] .
\label{eq:scatt_1D}
\end{eqnarray}
In this situation, the system  which  is frozen along the transverse
direction, can  be considered as quasi-1D. The even (odd) scattering
amplitude ${f^{\rm even}}$ (${f^{\rm  odd}}$) results from the asymptotic
contributions (${z\gg  a_\perp}$) in the different even (odd) 3D partial
waves. Quasi-1D scattering in the $s$-wave channel has been thoroughly studied
\cite{Maxim1D} and in the following we consider quasi-1D scattering of
two spin polarized  fermions in the vicinity of a $p$-wave resonance.
Thus, one can  restrict the source  term in Eq.(\ref{eq:Lippmann}) 
to the ${l=1}$ contribution, where ${{\mathcal S}_{1,{\mathbf  k}}={\bf k}/k}$, and the
choice ${|\Psi_0\rangle}$ imposes that ${{\mathcal R_1}= P_{1D} {\bf \hat{e}}_z}$. 
Computation of ${P_{1D}}$ from Eq.(\ref{eq:contact}) is much simpler 
in the domain of negative energy ${E<0}$ and using Eq.(\ref{eq:Lippmann}) 
one can show that:
\begin{equation}
P_{1D} = \frac{-3 q \phi_{00}(0) }
{\displaystyle \frac{1}{a_{1}(q)} + \frac{6}{\sqrt{\pi}a_\perp^3}
\mbox{P.f.}\! \int_0^\infty \!\! \frac{du}{u^{3/2}} \frac{\exp(\tau u)} {1-\exp(-u)}} .
\label{eq:dipole}
\end{equation}
where ${\tau=E/2\hbar \omega_\perp<0}$ and $\mbox{P.f.}$ denotes the ``partie finie de
Hadamard'' of the integral \cite{Partiefinie}. Interestingly, one recognizes 
the Riemann-Hurwitz Zeta function ${\zeta_H(-1/2,-\tau)}$ in the regularized integral 
of Eq.(\ref{eq:dipole}). Hence, by analytic continuation  
in  the domain ${\tau>0}$,  one  obtains a simple expression of  the scattering 
amplitude,  and in  the low energy limit ${q a_\perp \ll 1}$:
\begin{equation}
f_p^{\rm odd}=2i\pi P_{1D}\phi_{00}^*(0)
\simeq-iq\left(\frac{1}{l_p}+iq+q^2\xi_p\right)^{-1},
\label{eq:fpodd}
\end{equation}
with the odd-wave scattering length $l_p$ and the effective range $\xi_p$ 
given respectively by:
\begin{equation}
l_p= 6 a_\perp \left[ \frac{a_\perp^3}{w_1}-12\, \zeta(-\frac{1}{2}) \right]^{-1} \quad 
\mbox{and} \quad  \xi_p = \frac{\alpha_1 a_\perp^2}{6}.
\label{eq:lp}
\end{equation}
Expression of $l_p$ in Eq.(\ref{eq:lp}) coincides with the result
in Ref.\cite{Blume1D}, while the effective range $\xi_p$
and its physical implications was (up to our knowledge) not studied in previous works. 
However, $\xi_p$  is in general  not negligible for  all ${q\ll a_\perp^{-1}}$ and  crucially 
depends on the  width of the 3D resonance and on $a_\perp$. Indeed, the inequality 
${\xi_p\gg a_\perp}$ is likely to  occur as a consequence 
of Eq.(\ref{eq:inequality}) (${\alpha_1 R \gtrsim 1}$)  and  also  from the condition 
${a_\perp\gg R}$  which is needed from  the  hypothesis that  scattering processes 
are studied at collisional energies  ${\ll \hbar^2/\mu R^2}$. The regime 
${\xi_p \sim a_\perp}$ can be reached only for an extreme transverse confinement 
(${a_\perp \sim 10 R}$) and for the broadest $p$-wave resonances with ${\alpha_1 R \sim 1}$. 
In actual experiments, two $p$-wave resonances are used: the resonance for $^6$Li atoms 
({${R \simeq 3}$~nm}) at {$B_0\simeq215$~G} with ${\alpha_1 R \simeq 5}$ 
\cite{Chevy} and the one at {${B_0 \simeq 198.8}$~G} for $^{40}$K atoms {(${R\simeq 7}$~nm)} 
where ${\alpha_1 R \simeq 3}$ \cite{Ticknor}. For  ${l_p>0}$ the scattering amplitude 
in Eq.(\ref{eq:fpodd}) has a pole at ${q=i\kappa}$ with ${\kappa=(-1+\sqrt{1+4 \xi_p/l_p})/2\xi_p>0}$ 
giving a shallow bound state energy at (${-\hbar^2 \kappa^2/2\mu}$). For large and positive values of 
${l_p \gg\xi_p}$, ${\kappa=l_p^{-1}}$ and the bound state has a vanishing energy (\emph{i.e.} 
${\ll \hbar \omega_\perp}$) \cite{3D}. The scattering amplitude in Eq.(\ref{eq:fpodd}) can be 
obtained from  a 1D effective theory where the wave function $\psi_{1D}$ in 
Eq.(\ref{eq:scatt_1D}) solves the non interacting Schr\"{o}dinger equation and 
satisfies the following contact condition:
\begin{equation}
\lim_{z \to 0^+} \left(\frac{1}{l_p} + \partial_z - \xi_p \partial_z^2\right) \psi^{\rm odd}_{1D}(z)  = 0 ,
\label{eq:contact1Dodd}
\end{equation}
where $\psi^{\rm odd}_{1D}(z)= [\psi_{1D}(z)-\psi_{1D}(-z)]/2$ is the projection of the wave 
function onto its odd component \cite{Pseudo}. This approach can be generalized 
for few- and many-body systems by imposing the contact condition in 
Eq.(\ref{eq:contact1Dodd}) for each pair of interacting particles. Without performing 
these calculations which are clearly beyond the scope of this letter, it is of 
importance to determine the conditions such that the fully polarized fermionic gas 
can reach the FTG regime or equivalently  can  be mapped onto non interacting 1D 
bosons \cite{Wright}. Eq.(\ref{eq:contact1Dodd}) implies that this mapping is 
possible only at resonance ($|l_p|=\infty$) and also if the effective range is negligible. 
The resonant condition is easily obtained by using a Feshbach resonance. However, 
the momentum distribution of the FTG gas has a large tail given by a Lorentzian of 
width $4n$, where $n$ is the 1D atomic density \cite{Bender}, thus $\xi_p$ can be 
neglected only in the dilute limit: $n \xi_p \ll 1 $.
For $N$ atoms trapped in a strongly anisotropic trap with a weak harmonic 
confinement along the $z$-direction (atomic frequency ${\omega_z\ll\omega_\perp}$ and axial 
length ${a_z=\sqrt{\hbar/\mu\omega_z} \gg a_\perp}$), in the FTG regime one has ${n \sim N/a_z}$ 
and the condition $n \xi_p \ll 1 $ gives ${N\ll6a_z/(\alpha_1  a_\perp^2)}$. 
For example, considering $^{40}K$ atoms in a highly anisotropic trap with 
${\omega_\perp=2\pi \times70}$~kHz and ${\omega_z=2\pi \times  10}$~Hz, the FTG regime is obtained 
for ${N \ll 14}$, which makes sense only for few-body configurations 
(for $^6$Li atoms with the same trap parameters, 
$\xi_p$ is of the order of $a_z$ and the mapping to a non interacting Bose system is  
a poor approximation even for the two-body ground state).
Interestingly, if the condition $n \xi_p \ll 1 $ is not satisfied, the scale 
invariance in the linear wave guide is broken at low energy, hence 
the  time-dependent many-body ansatz in Ref.\cite{AnnaFTG} is no more 
an exact solution. To conclude this part, excepted specific configurations, 
the criterium $n \xi_p \ll 1 $ is in general not verified and the density corrections 
to the FTG properties are important issues for understanding the resonant gas. 
Moreover, following the reasoning of the box model in Refs.\cite{pwave,Box}, one expects 
that by varying $l_p$ from large and positive values to large and  negative values, 
the system experiences a narrow $p$-wave BEC-BCS cross-over, where the composite 
bosons in the dilute BEC phase (${l_p>0}$) are dimers populating the quasi-1D 
shallow two-body bound state. 

{\it Planar atomic wave guide}. --- High partial wave superfluidity in
quasi-2D geometries is interesting for its links with condensed matter
physics like for example high-$T_c$ superconductivity and the possible 
applications for quantum computing \cite{Gurarie,Tewari}. These studies 
motivate  a close investigation of quasi-2D scattering properties. 
In this geometry, the two colliding atoms are confined in a planar 
harmonic trap along the $z$-direction and move freely in the ${(xy)}$ 
plane. The free Hamiltonian reads:
\begin{equation}
{\mathcal H}_0 = - \frac{\hbar^2}{2\mu} \Delta_{\bf r} + \frac{1}{2} m \omega_z^2 z^2 
-\frac{\hbar\omega_z}{2} .
\label{eq:H2D}
\end{equation}
The homogeneous solution corresponding to a scattering process at energy
${E=\hbar^2q^2/2\mu}$ is: $\langle{\bf k}|\Psi_0\rangle = (2\pi)^2 \delta({\mathbf k}_{2D}-{\mathbf q})\,\langle k_z|\phi_0\rangle$,  
where ${|\phi_0\rangle}$ is the ground state of the 1D harmonic oscillator in
Eq.(\ref{eq:H2D}). In the monomode regime (${E < \hbar\omega_z}$), the system can 
be considered as quasi-2D, and for large interparticle separation (${\rho \gg
a_z}$), the wave function factorizes: ${\Psi({\mathbf  r}) = \phi_0(z) \psi_{2D}({\mathbf \rho})}$. The 2D partial scattering amplitudes ${f^{[m]}=f^{[-m]}}$ 
can be defined by the following expansion:
\begin{equation}
\psi_{2D}({\boldsymbol \rho}) \operatornamewithlimits{=}_{\rho \gg a_z} e^{(i{\mathbf q}.{\boldsymbol \rho})} 
- \frac{i}{4} \sum_{m=-\infty}^{m=\infty}  f^{[m]}  H_{m}^{(1)}(q\rho)  e^{(im\theta)} ,
\label{eq:2Dscatt}
\end{equation}
where  ${\theta = \pi/2 + \angle {({\boldsymbol \rho},{\mathbf  q})}}$  and
${H_{m}^{(1)}}$ is the Hankel's function. Using  
Eqs.(\ref{eq:Lippmann},\ref{eq:contact},\ref{eq:2Dscatt}) the  scattering 
amplitude for two atoms interacting in the $s$-wave channel is:
${f^{[0]}= 4\pi [a_z\sqrt{\pi}/a_0(q)+J_0(\tau+i0^+)]^{-1}} $, with ${\tau=E/2\hbar\omega_z}$, 
$J_0(\tau)=\ln(-B/(2\pi \tau))+\sum_{n=1}^\infty \ln(n/(n-\tau)) (2n-1)!!/(2n)!!$ \cite{Jm},
and ${B \simeq  0.9049}$ \cite{constante_B}. In the  $p$-wave channel, Eq.(\ref{eq:inequality}) 
implies the existence of a  small parameter ${\eta_1=(\alpha_1a_z)^{-1}\ll 1}$ which plays a central 
role in the scattering properties \cite{Example}. For 
${{\mathbf q} = q {\bf \hat{e}}_x}$, the $p$-wave regular part in Eq.(\ref{eq:Lippmann}) 
is related to the 2D $p$-wave scattering amplitude $f^{[1]}$ by 
${\mathcal R}_1= - f^{[1]}(q){\bf \hat{e}}_x /[2 \pi q \phi_0^*(0)]$. Using the contact condition in 
Eq.(\ref{eq:contact}) where ${{\mathcal S}_{1,{\mathbf k}}={\bf k}/k}$ gives:
\begin{equation}
f^{[1]}(q) = 6 \pi q^2 |\phi_0(0)|^2 \left[\frac{1}{a_1(q)}+\frac{6 J_1(\tau+ i 0^+)}{\sqrt{\pi} a_z^3} \right]^{-1} ,
\label{eq:f1}
\end{equation}
where  ${|\phi_0(0)|^2=1/(a_z\sqrt{\pi})}$. In the low energy limit ${|\tau|\to0}$ then $J_1(\tau)=J_1(0)+\tau\ln(-eB/2\pi\tau)+O(\tau^2)$,   
where $J_1(0)\simeq-5.4722\times10^{-2}$ \cite{Jm}, and the logarithmic term ensures
the unitarity condition in Eq.(\ref{eq:f1}). A similar calculation can be done for atoms 
interacting resonantly in the 3D $d$-wave channel. In this case, the small parameter related 
to the resonance width is ${\eta_2=\alpha_2^{-1}a_z^{-3}\ll 1}$. For ${{\bf q}=q {\bf \hat e}_x}$, 
the $l=2$ regular part is given by: $k^2 \left(  {\mathcal R_2}  \!\cdot\! {\mathcal S}_{2,{\bf  k}} \right) 
= S_{2D} (2 k_z^2-k_{2D}^2)+D_{2D} (k_x^2-k_y^2)$, where $S_{2D}$ contributes to the ${m=0}$ 2D partial 
wave channel and $D_{2D}$ contributes to the ${|m|=2}$ channel (2D-$d$-wave). The scattering 
amplitude in the ${|m|=2}$ channel is ${f^{[2]}(q) = - 2\pi q^2\phi_0^*(0)D_{2D}}$, and using Eq.(\ref{eq:contact}): 
\begin{equation}
f^{[2]}(q) = \frac{15 \pi q^4 |\phi_0(0)|^2}{2} \left[\frac{1}{a_2(q)} + \frac{60 J_2(\tau+ i 0^+)}{a_z^5\sqrt{\pi}} \right]^{-1} ,
\label{eq:f2}
\end{equation}
where ${J_2(\tau) =  J_2(0) + \tau J_1(0)  + O(\tau^2)}$ for ${\tau \to  0}$, and ${J_2(0)\simeq -2.2752\times 10^{-2}}$ 
\cite{Jm}.   
From Eqs.(\ref{eq:f1},\ref{eq:f2}), several conclusions can be drawn on the quasi-2D scattering 
in ${m=1}$ and ${m=2}$ partial wave channels: i) At small collisional energy (${q \to 0}$), 
${f^{[m]}(q) \propto q^{2m}/(w_m^{-1}-w_m^{\star-1})}$  where ${w_1^\star\simeq5.39 \times a_z^3}$ and $w_2^\star\simeq1.3\times a_z^5$. Hence, the 2D 
confinement induces a shift $w_m^\star$ in the generalized scattering length $w_m$, which grows as the 
confinement increases. ii) By modifying the trap frequency and/or the external magnetic field it 
is possible to drive the system from a regime with a 2D $m$-wave quasi-bound state (${w_m^{-1}<w_m^{\star-1}}$) 
to a regime with a shallow bound state (${0<w_m\lesssim w_m^\star}$), in both cases the (quasi-) bound state 
energy is ${E_b\sim-\hbar^2(w_m^{-1}-w_m^{\star-1})/(2\mu\alpha_m)}$  \cite{criterium}, thus having an expression similar 
to the 3D case: ${E_b^{3D}\sim-\hbar^2/(2\mu w_l\alpha_l)}$. iii) Resonance in the scattering cross section ${\sigma=|f_m|^2/4q}$ 
occurs at a collisional energy ${E\sim E_b}$, that is only for {\sl positive} values of $E_b$: in presence 
of a quasi-bound state. The resonance width is given by ${\Delta E/E_b \propto \eta_m (E_b/\hbar\omega_z)^{m-1} \ll 1}$.
iv) Consequently,  $\alpha_l$  appears as a crucial parameter for the low energy scattering properties. For example, 
by neglecting $\alpha_1$ (${\eta_1=\infty}$) in Ref.\cite{Idziaszek}, the resonant collisional energy 
and resonance width of the $p$-wave CIR are found with other order of magnitudes. 
v) Following the reasoning in Ref.\cite{Box}, these scattering properties open the possibility of observing 
high partial waves BEC-BCS transitions in quasi-2D fermionic gases by varying ${(w_m^{-1}-w_m^{\star-1})}$ 
from positive to negative values.

{\bf Acknowledgments}

It is a pleasure thank Y. Castin, M. Holzmann, M. Jona Lasinio, 
A. Minguzzi, P. Pedri and  F. Werner for discussions. {\it LPTMC} 
is {\it UMR  7600 of CNRS}.


\begin{thebibliography}{99}
\bibitem{deltalike} This delta-like function has been introduced
implicitly for the usual $s$-wave zero-range potential in Ref.\cite{reseau}.
\bibitem{3D} In the regime where ${R^3 \ll w_1 \ll a_\perp^3}$, then ${l_p\ll   a_\perp\ll\xi_p}$ and ${\kappa \simeq  1/\sqrt{\alpha_1 w_1}}$
which coincides with the 3D result.
\bibitem{Pseudo}  Eq.(\ref{eq:contact1Dodd}) can be obtained also by use of a $\Lambda$-potential:
$
\langle z |V_\Lambda|\phi \rangle =\displaystyle  -\hbar^2 l_p \delta'(z)
\lim_{\epsilon \to 0^+} (\Lambda  + \partial_\epsilon - \xi_p \partial_\epsilon^2 ) \phi^{\rm odd}(\epsilon)/[\mu (1 -\Lambda  l_p)]
$
, where $\Lambda$ is a free parameter \cite{lambdapot,lwaves,pwave}.
\bibitem{Jm} The functions $J_m(\tau)$ are  defined in the domain ${\tau<0}$ by: 
$
J_m(\tau) = \mbox{P.f.}\! \int_0^\infty \!\! du\, {u^{-(m+1)}} \exp(\tau u) [1-\exp(-u)]^{-1/2} 
$,
and appear in the computation of $f^{[m]}$. The expression  of $J_m(\tau+i0^+)$ 
are deduced using  an analytical continuation.
\bibitem{constante_B} The constant $B$ is obtained from:
$
\ln\left(B/2\pi\right) = -\gamma + \int_0^\infty \!\! du\, \left\{ u^{-1} [1-\exp(-u)]^{-1/2}-u^{-3/2}-(1+u)^{-1}  \right\}
$, 
where  $\gamma$ is  the Euler's  constant. This value slightly differs from 
the one in Ref.\cite{Petrov2D}. However, a close investigation shows 
that it coincides with the evaluation of the slowly converging series 
in Ref.\cite{Petrov2D}.
\bibitem{criterium} For ${m=1}$, this expression of  $E_b$ neglects the logarithmic 
correction which is vanishingly small if ${\eta_1 \left|\ln\left| a_z^3(w_1^{-1} -w_1^{\star-1}) \right| \right| \ll 1}$.
\end{thebibliography}
\end{document}